\documentstyle[prl,aps,preprint,tighten,psfig,amsmath,ifthen]{revtex}
\draft

\newcommand{\xor}{\oplus}

\newcommand{\eq}[1]{
	\begin{equation*}
		#1
	\end{equation*}
}
\newcommand{\eql}[2]{
	\begin{equation}
		\label{eq:#1}
		#2 
	\end{equation}
}
\newcommand{\meq}[1]{
	\begin{equation*}
		\begin{split}
			#1
		\end{split}
	\end{equation*}
}

\renewcommand{\eqref}[1]{%
(\protect\ref{eq:#1})%
}

\begin{document}

\title{Synchronization and Maximum Lyapunov Exponents of Cellular
Automata}

\author{Franco Bagnoli$^{(1)}$~\cite{email} and Ra\'ul Rechtman$^{(2)}$}

\address{
$(1)$ Dipartimento di Matematica Applicata, Universit\`a di Firenze,
Via S. Marta, 3 I-50139 Firenze, Italy. \\Also INFN and INFM, sez. di Firenze, Italy\\
$(2)$ Centro de Investigac\'{\i}on en Energ\'{\i}a,
Universidad Nacional Aut\'onoma de M\'exico, \\
62580 Temixco, Morelos, Mexico}

\date{today}

\maketitle

\begin{abstract}
	We study the synchronization of totalistic one dimensional
	cellular automata (CA). The CA with a non zero synchronization 
	threshold exhibit
	complex non periodic space time patterns and conversely. This
	synchronization transition is related to directed percolation.
    We study also the maximum Lyapunov exponent for CA, defined
	in analogy with continuous dynamical systems as the
	exponential rate of expansion of the linear map induced by the
	evolution rule of CA, constructed with the aid of the Boolean derivatives.	
	The synchronization threshold is strongly correlated to 
	the maximum Lyapunov exponent and we propose approximate
	relations between these quantities.  The value of this threshold 
	can be used to parametrize the space time
	complexity of CA. 
\end{abstract}



Cellular automata (CA) are discrete dynamical systems that may
exhibit complex space time patterns~\cite{wol,wol84}. 
It has been observed that CA may be synchronized by a stochastic 
coupling~\cite{bag95,zan98,gra98}.  We study all totalistic CA with
four, five and six neighbors.  
For the CA considered, we find
that a synchronization threshold is reached critically and that
all CA with complex non periodic space time patterns have a positive
threshold and conversely. We also find a strong 
relationship between the synchronization threshold and the maximum
Lyapunov exponent (MLE) of CA~\cite{bag92,others}. 

Let us start considering the following asymmetric coupling
for a continuous one-dimensional map $f(x)$~\cite{pik91}
\meq{
		x' &= f(x),\\ 
		y' &= (1-p)f(y) + p f(x),
}
with $x=x(t)$, $x^\prime=x(t+1)$ (idem for $y$) and $0\leq p\leq 1$.
The function $f(x)$ depends in general by one (or more) parameter $a$; 
let assume that, for the chosen value of $a$, $f(x)$ is chaotic 
with Lyapunov exponent $\lambda$, and 
that $x(0) \ne y(0)$.  Then,  $x(t)$ is always different from $y(t)$ 
for $p=0$, while for $p=1$  $x$ and $y$ synchronize in one time step.
There exists a critical synchronization threshold $p_c$ for which both
trajectories $x(t)$ and  $y(t)$ become
indistinguishable in the long time limit and
\eql{esmap}{
	p_c = 1-\exp(-\lambda).
}

In what follow we shall try to develop similar relations for CA.
We begin with a brief review of the definition of maximum 
Lyapunov exponent for CA based on a
linear expansion of  the evolution rule. We then present a
synchronization  mechanism and show that the distance between
two realizations goes to zero in a critical manner at
$p_c$. The numerical experiments show a relation
between $p_c$ and the maximum Lyapunov
exponent which may be understood by considering several probabilistic
CA.  We restrict our study to one dimensional,
totalistic Boolean cellular automata with four, five and six inputs, since their 
number is reasonably manageable and their evolution can be efficiently
implemented~\cite{bag92a}.

A Boolean CA $F$ of range $r$ is defined as a map on the set of 
configurations $\{{\bf x}\}$ with ${\bf x} = (x_0,\dots ,x_{N-1})$, 
$x_i =0,1$, and $i=0,\dots ,N-1$ such that
\[
  {\bf x}' = F({\bf x})
\]
where ${\bf x}={\bf x}(t)$, 
${\bf x}'={\bf x}(t+1)$ and $t=0,1\dots$. 
The map $F$ is defined locally on every site $i$ by
\eq{
 x_i' = f(\{x_{i}\}_r)
}
where $\{x_{i}\}_r=(x_{i},\dots ,x_{i+r-1})$ is the neighborhood of
range $r$ of site $i$ at time $t$, assuming periodic boundary conditions. 
For totalistic CA, the local function 
$f$ is symmetric and depends only on $s$ defined by 
\eq{
 s(\{x_i\}_r) = \sum_{j=0}^{r-1} x_{i+j}.
}
That is $x_i' = f(s(\{x_i\}_r))$.
It is useful to introduce the following operations between Boolean
quantities: the sum modulo two (XOR), denoted by the symbol $\xor$,
and the AND operation, which is analogous to the usual multiplication
and shares the same symbol. These operations can be performed between two
configurations component by component. 
We introduce the difference, or damage,  $\bf z=x\xor y$,
whose evolution is given by ${\bf z}'=F({\bf x})\oplus F({\bf y})$
and we define the norm of ${\bf z}$ as $|{\bf z}|=(1/N)\sum_i
x_i\oplus y_i$. 

A function $f(x_i, \dots, x_j, \dots, x_{i+r})$ is sensitive to its
$j$-th argument for a given neighborhood $(\{x_i\}_r)$ 
if the Boolean derivative 
\eq{
 \left.\frac{\partial f}{\partial x_j}\right|_{\{x_i\}_r} =
  f(x_i,\dots, x_j, \dots)\oplus 
  f(x_i,\dots, x_j\xor 1, \dots)
}
is 1. The Jacobian matrix $J$ of $F$ is an $N\times N$ matrix with components 
\eq{
  J_{i,j}({\bf x}) = \left.\frac{\partial f}{\partial x_j}
  	\right|_{\{x_i\}_r}.
}
The matrix $J$ is circular with zeroes everywhere except possibly on the main
diagonal and the following $r-1$ upper diagonals.
 
It is possible to ``Taylor expand'' a Boolean function around a
given point using Boolean derivatives~\cite{Bagnoli:Minimization}. 
To first order in $|\bf z|$  we have
\eql{efirst}{
 F({\bf y})=F({\bf x})\oplus J({\bf x})\odot{\bf z}
}
where  $\odot$ denotes the Boolean 
multiplication of a matrix by a vector. Compared to algebraic multiplication of
a matrix by a vector, the sum and multiplication of scalars are replaced by the
XOR and the AND operations respectively. 
Using Eq.~\eqref{efirst} we may approximate the evolution of the damage
configuration ${\bf z}$ by
\eq{
  {\bf z}'=J({\bf x})\odot {\bf z}.
}
However, $|{\bf z}|$ grows at most linearly with $t$ since a damage cannot 
spread to more than $r$ neighbors in one time step: a fixed site $i$ at time 
$t+1$ can be damaged if at least one of its $r$ neighbors at time $t$ is 
damaged, but if more than one of the neighbors is damaged, the damage may
cancel. Since
\eq{
 z_i' = \bigoplus_{j=i}^{i+r-1} J_{i,j}({\bf x}) z_j,
}
$z_i'=1$ if $J_{i,j}({\bf x}) z_j=1$ on an odd
number of sites. In order to account for all possible damage spreading we 
choose to consider each damage independently. If, at time $t$, 
$m$ damaged sites
are present, we consider $m$ replicas each one with a different damaged site. 
On each replica, the damage evolves for one time step, without interference
effects and so on.

This procedure is equivalent to choosing a vector
$\mbox{\boldmath $\xi$}(0)={\bf z}(0)$ 
that evolves in time according to 
\eql{etangent}{
  \mbox{\boldmath $\xi$}'=J({\bf x})\mbox{\boldmath $\xi$}
}
where now the matrix multiplication is algebraic. The components ${\xi_i}$ 
are positive integers that count the number of ways in which the initial
damage can spread to site $i$ at time $t$ on the ensemble of replicas.

\begin{figure}[t]
 \centerline{\psfig{figure=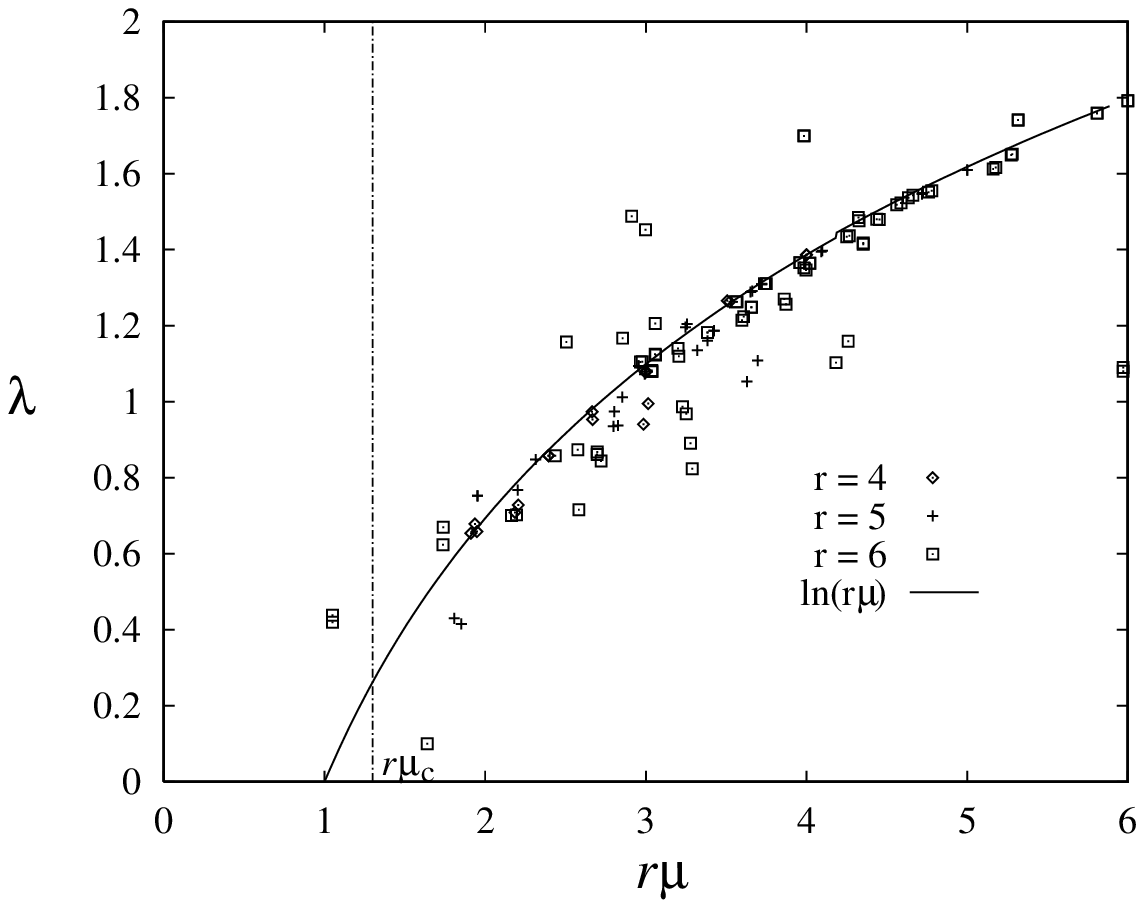,height=10cm,width=12cm}}
 \begin{center}
  \begin{minipage}{8.5cm}
   \caption{\label{flam-mu} The MLE of totalistic CA with $r=4,5,6$
    versus $r\mu$. The
    continuous line represents the mean field approximation 
    $\tilde \lambda = \ln(r\mu)$. The dashed line marks the threshold
    $r\mu_c$.}
  \end{minipage} 
 \end{center}
\end{figure}

We define the maximum Lyapunov exponent $\lambda$ of the cellular automaton 
$F$ by
\eq{
 \lambda({\bf x}^0) = \lim_{T\rightarrow\infty}\lim_{N\rightarrow\infty}
           \frac{1}{T}\log\left(\frac{|\mbox{\boldmath $\xi$}^T|}
                                     {|\mbox{\boldmath $\xi$}^0|}\right),
}
where $|\mbox{\boldmath $\xi$}|$ may be taken as the Euclidean norm
or as the sum of its components. The geometrical
average $\mu$ of ones in the  Jacobian matrix $J$ is defined by 
\eq{
 \mu({\bf x}^0) = \lim_{T\rightarrow\infty}\lim_{N\rightarrow\infty}
 \left(\prod_{t=0}^{T-1}\frac{1}{r}\sum_{i,j} J_{i,j}({\bf x}^t)\right)^{1/T}.
} 
We show in Fig.~\ref{flam-mu} the points $(r \mu,\lambda)$ of all the
totalistic CA  with $r=4,5$, and 6 that show complex non periodic
space time patterns.

The process defined by Eq.~\eqref{etangent} may be viewed as a
deterministic directed bond  percolation problem where a site $i$ at
depth $t$ is wet if $\xi_i(t)>0$. The bonds exist where the
components of $J$ are 1.  A first approximation is obtained by
replacing $J$ with a random matrix whose elements are zero except on
the diagonal and the $r-1$ following upper diagonals, where they  are
equal to one with probability  $\mu$~\cite{bag92,bag93,bag94}. There
is a critical value $\mu_c(r)$  below which the bond percolation
process falls into the absorbing state so that the maximum eigenvalue
of the product of random matrices is zero.  We can further introduce
a mean field approximation to the directed bond percolation process,
which exhibits discrepancies only very near to $\mu_c$. In this case
one can show that
 \eql{elrandom}{
 \tilde \lambda = \ln(r\mu)
}
is an upper bound to the MLE of the product of random matrices. 
The behavior of  $\tilde \lambda$ is plotted in 
Fig.~\ref{flam-mu}, and we note that it is a
good approximation to the generic behavior of CA. We also report 
the value $r\mu_c(r)$ for which the maximum eigenvalue of the product of random
matrices is zero, corresponding to the percolation threshold for the
directed bond percolation problem. We found that $r \mu_c(r)\simeq 1.3$ regardless
of $r$, a fact that can be understood from the following
argument. The percolation cluster has on the average $\kappa$ bonds per site with 
$\kappa(r) = r\mu_c(r)$; for this percolation model, one connection to a wet
neighbor at the previous time step  is sufficient, and further connections do
not alter wetting. 

We now discuss a synchronization mechanism for CA. Starting with
two initial configurations chosen at random ${\bf x}(0)$ and ${\bf
y}(0)$ we propose that
\meq{
 {\bf x}'&=F({\bf x}),  \\
 {\bf y}'&=\overline {S^t(p)}F({\bf y})\oplus 
 					S^t(p)F({\bf x}), 
}
where $S^t(p)$ is a Boolean random diagonal matrix with elements $s^t_i(p)$
that, at each time step, take the value one with probability $p$ 
and zero with probability $1-p$;
$\overline {S(p)} = I-S(p)$ and $I$ is the identity matrix. On the average,
$y_i'$ will be set to the value of $x_i'= f(\{x_i\})$ on a 
fraction $p$ of sites. 
In this way we introduced a stochastic synchronization mechanism over a
deterministic process. This stochatic 
mechanism can be considered as a ``random field''
approximation of an intermittent coupling generated by a deterministic chaotic
process.
The evolution equation for the difference $\bf z=x\oplus y$ is 
\eql{z}{
 {\bf z}'=\overline {S(p)}\left[F({\bf x})\oplus F({\bf y})\right].
}
The control and order parameters are $p$ and 
$h(p)=\lim_{t\rightarrow\infty}\lim_{N\rightarrow\infty}|{\bf z}(t)|$ respectively. 
We say that ${\bf x}$, the driver, and ${\bf y}$, the driven
system, synchronize when $h(p) = 0$.  For $p=0$ both
systems evolve independently, while for $p=1$ they synchronize in just
one step; we expect then to find a synchronization threshold $p_c$.
This behavior is shared by all the CA with complex non periodic space
time patterns.  All others synchronize for  $p\simeq 0$. This can be
conversely expressed by saying that all CA that  synchronize with a
nontrivial $p_c$ exhibit complex non periodic space time patterns. 

For totalistic linear rules, whose evolution is given by
\eq{
	f(\{x_i\}_r) =\bigoplus_{j=0}^{r-1} x_{i+j},
}
the synchronization equation \eqref{z} is equivalent to the dilution
(with probability $1-p$) of the rule. For $r=2$ the dilution problem
is equivalent to the line $z=0$ of Ref.~\cite{DK}, whose transition 
belongs to the universality class of directed percolation (DP)~\cite{tome}.
For a similar synchronization mechanism, it has been recently claimed
that the elementary CA rule 18 does not~\cite{zan98} and does~\cite{gra98}
belong to the DP universality class. The presence of a single
absorbing state and the absence of other conserved quantities (i.e.\
number of kinks) strongly suggests that the synchronization transition
belongs to the DP universality class~\cite{bag95,gra98}.
The goal of this work was not that of computing critical exponents. However,
since the critical point was located by means of the scaling law for
the density of defects (in order to minimize finite size and time effects), 
we had the opportunity of computing the
magnetic exponent $\beta$ for all the CA studied. Due to the large
number of CA examined, this computation was
performed semi-automatically, and the precision of the resulting
exponent is quite low, nontheless 
we have not found any example of non-DP behavior. We performed bulk 
simulations for $N=2000$ and $T=4000$ and checked that the results do not 
change for doubling of halving these figures.

\begin{figure}[t]
 \centerline{\psfig{figure=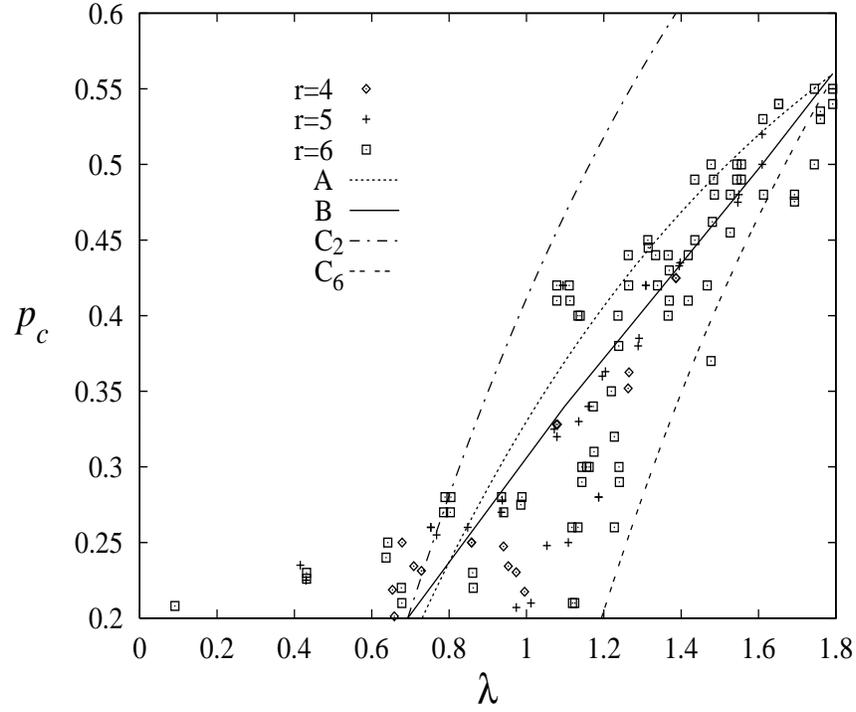,height=10cm,width=12cm}}
 \begin{center}
  \begin{minipage}{8.5cm}
   \caption{\label{fpc_lam} Relationship between $p_c$ and $\lambda$ for
    all CA with range $r=4,5,6$ (markers) and complex space time patterns. 
    The curves correspond to the various approximations, as specified in the legend.}
  \end{minipage} 
 \end{center}
\end{figure}

In Fig.~\ref{fpc_lam} we 
plot the points $(p_c,\lambda)$ for all totalistic CA with $r=4,5,6$
and a non trivial value of $p_c$.  All the numerical experiments were performed
using a parallel algorithm that takes care of all the values of $p$ 
simultaneously~\cite{bag97}. 

Let us study the relation between $p_c$ and $\lambda$ by some random
approximations. 
Near the synchronization threshold $p_c$, we may expand ${\bf y}$ around
${\bf x}$ with the help of Eq.~\eqref{efirst} so that
\eql{esinc}{
 {\bf z}'=\overline {S(p)}J({\bf x}){\bf z}.
}	
Eq.~\eqref{esinc} may be written as
\eql{esinci}{
 z_i'=\overline {s_i(p)}
 		\left[J_{i,i}({\bf x})z_i\oplus\cdots\oplus
 		J_{i,i+r-1}({\bf x}) z_{i+r-1}\right].
}
During the time evolution of a particular CA a fixed value of $\mu$
is attained so that  on average $r\mu$ derivatives inside the
parenthesis on the {\em r.h.s.} of Eq.~\eqref{esinci}  are different
from zero. A first approximation, model {\em A}, is  obtained by
replacing the derivatives with random variables  $m_i(\mu)$ that are
one with probability $\mu$ and zero with probability 
($1-\mu$). That is
\eq{
 z_i'=\overline {s_i(p)}
 			\left[m_i(\mu)z_i\oplus\cdots\oplus
 			m_{i+r-1}(\mu)z_{i+r-1}\right].
}
On every site and every time step, 
$r\mu$ variables are chosen on average and if their sum is odd, then with 
probability $p$, $z_i'=1$. We then look for the synchronization 
threshold $p_c(\mu)$ and plot it as a function of $\tilde\lambda=\ln(r\mu)$,
Eq.~\eqref{elrandom}. This is the curve labeled $A$ in Fig.~\ref{fpc_lam}. 
The predicted values of $p_c$ are generally higher than those found for CA
for the same value of $\lambda$ possibly due to correlations
among the derivatives.

Since the typical ``complex'' CA pattern exhibit transient
correlations (``triangles''), 
one can model them by choosing a fixed number $k\leq r$ of derivatives 
equal to one. The simplest way is to take $k=r\mu$ with $r\mu$ an integer,
model {\em B}, which . Then
\eq{
 z_i'=\overline {s_i(p)}
 			\left[z_i\oplus\cdots\oplus z_{i+k-1}\right].
}
which is a dilution of the XOR with $k$ inputs. This process is
expected to belong to the same universality class of directed
percolation~\cite{Grassberger:conjecture}. The curve labeled {\em B}
of Fig.~\ref{fpc_lam} passes through all the calculated values for
$k=2,\dots 6$. We note that this second model is a better fit of CA behavior.

We can extend this last model allowing noninteger values of $r\mu$ by
\eq{
 z_i(t+1)=\overline {s_i^t(p)}m_i^t(\mu)
 			\left[z_i\oplus\cdots\oplus z_{i+k-1}\right]
}
which we call model {\em C$_k$}. 
In this way the average number of non zero derivatives is
$k\mu$ with $0\leq \mu \leq 1$.  
Now $k$ is a free parameter, and this
model can be useful to delimit the expected spread of $(\lambda,p_c)$
points. 

Since $s_i^t(p)$ and $m_i^t(\mu)$ are independent random variables we may
write
\eq{
 z_i(t+1)=\overline {s_i^t(q)}
  			\left[z_i\oplus\cdots\oplus z_{i+k-1}\right].
}
where $(1-q)=(1-p)\mu$. In this guise, this is model {\em B} with $k$
inputs.
The synchronization threshold is given by $p_c(k,\mu)=1-(1-q_c(k))/\mu$ 
where $q_c(k)$ is
the percolation threshold of the dilution of the XOR with $k$ inputs. 
Using the approximation
$\lambda=\ln(k\mu)$ one has
\eql{last}{ 
 p_c(k,\lambda)=1-k\bigl(1-q_c(k)\bigr)\exp(-\lambda).
}
which bears some likeness to Eq.~\eqref{esmap}.
The curves labeled $C_2$ and $C_6$ of Fig.~\ref{fpc_lam} correspond to 
this last expression for $k=2$ and $k=6$ respectively. 
We note that the points $(\lambda,p_c)$ of 
almost all the CA considered fall between these two curves. 

For the totalistic one dimensional CA with $r=4,5,6$ we can safely say that
all CA with a positive $p_c$ exhibit complex non periodic space time patterns 
and conversely. These CA also have a positive MLE . We also showed that the
synchronization of CA is a critical phenomenon similar to directed percolation.

We proposed several approximations based on a combination of ``linearization''
of CA rules using Boolean Taylor expansions and stochasticity and showed the
relation between the synchronization threshold and the MLE. In particular,
model $C$ implies a relation similar to that found for continuous maps with the
addition of a percolation threshold constant. 

An analogous mechanism can be applied to coupled map lattices; in this case $p$
is the probability that a map $y_i(t)$ takes the value $x_i(t)$. One observes a
synchronization transitions, but the critical value $p_c$ is not correlated to
the usual MLE~\cite{bag98}.

\vspace{.3cm}
We acknowledge partial financial support from project DGAPA-UNAM IN103595.


\end{document}